\documentclass[]{raa}
\usepackage{amssymb, amsmath, epsf, graphicx}
\usepackage{natbib}

\catcode`\@=11 
\def\@versim#1#2{\vcenter{\offinterlineskip
    \ialign{$\m@th#1\hfil##\hfil$\crcr#2\crcr\sim\crcr } }}
\def\lsim{\mathrel{\mathpalette\@versim<}}
\def\gsim{\mathrel{\mathpalette\@versim>}}

\newcommand{\hei}{He {\sc I}}
\newcommand{\hi}{H {\sc I}}
\newcommand{\hii}{H {\sc II}}
\newcommand{\heii}{He {\sc II}}
\newcommand{\mgii}{Mg {\sc II}}
\newcommand{\caii}{Ca {\sc II}}
\newcommand{\nv}{N {\sc V}}
\newcommand{\siiv}{Si {\sc IV}}
\newcommand{\civ}{C {\sc IV}}
\newcommand{\feii}{Fe {\sc II}}
\newcommand{\feiii}{Fe {\sc III}}
\newcommand{\alii}{Al {\sc II}}
\newcommand{\aliii}{Al {\sc III}}
\begin{document}
\bibliographystyle{raa}
\title{Discovery of Balmer Broad Absorption Lines in the Quasar LBQS
  1206+1052}
\volnopage{ {\bf 20xx} Vol.\ {\bf 9} No. {\bf XX}, 000--000}
   \setcounter{page}{1}

   \author{Tuo Ji
      \inst{1,2}
   \and Tinggui Wang
      \inst{1,2}
   \and Hongyan Zhou
      \inst{1,2,3}
     \and Huiyuan Wang
     \inst{1,2}
   }
 \institute{Key Laboratory for Research in Galaxies and
   Cosmology, The University of Sciences and Technology of China
   (USTC), Chinese Academy of Sciences, Hefei, Anhui 230026, China;
   ~tji@mail.ustc.edu.cn
   \and
   Center for Astrophysics,USTC, Hefei, Anhui 230026, China
     \and
    Polar Research
    Institute of China,451 Jinqiao Road, Pudong, Shanghai 200136, China
     \vs \no
   {\small Received [2011] [10] [15]; accepted [2012] [01] [03] }
}


\abstract{
We report the discovery of Balmer broad absorption lines (BALs) in
the quasar LBQS 1206+1052 and present a detailed analysis of the
peculiar absorption line spectrum. Besides Mg~II~$\lambda \lambda
2796,~2803$ doublet, BALs are also detected in He~I* multiplet at
$\lambda \lambda 2946,~3189,~3889$~\AA~arising from metastable helium
$2^3S$ level, and in H$\alpha$ and H$\beta$ from excited hydrogen
H~I* $n=2$ level, which are rarely seen in quasar spectra. We identify
two components in the BAL troughs of $\Delta v\sim$2000~km~s$^{-1}$
width: One component shows an identical profile in H~I*, \hei* and
\mgii~with its centroid blueshifted by $-v_{\rm c}\approx 726$~km\
s$^{-1}$. The other component is detected in \hei* and \mgii~with
$-v_{\rm c}\approx 1412$~km~s$^{-1}$. We estimate the column densities of
H~I*, He~I*, and Mg~II, and compare them with possible level
population mechanisms. Our results favor the scenario that the
Balmer BALs originate in a partially ionized region with a column
density of $N_{\rm H}\sim 10^{21-22}$~cm$^{-2}$ for an electron density of
$n_e\sim 10^{6-8}~$cm$^{-3}$ via Ly$\alpha$ resonant scattering
pumping. The harsh conditions needed may help to explain the rarity
of Balmer absorption line systems in quasar spectra. With an
$i$-band PSF magnitude of 16.50, LBQS 1206+1052 is the brightest
Balmer-BAL quasar ever reported. Its high brightness and unique
spectral properties make LBQS 1206+1052 a promising candidate for
follow-up high-resolution spectroscopy, multi-band observations, and
long-term monitoring.
}
 \authorrunning{T. Ji, T.G. Wang, H.Y. Zhou  \& H.Y. Wang}
 \titlerunning{Balmer BAL LBQS  1206+1052}
 \maketitle

\section{Introduction}

Outflows from active galactic nuclei (AGN) play an important role in
galaxy evolution. Gas in AGN may be accelerated to high velocities
by thermal pressure \citep{1983ApJ...271...70B}, radiation pressure
due to line and continuum absorption \citep{2004ApJ...616..688P},
magneto-centrifugal forces \citep{2005ApJ...631..689E}, or their
combinations. The outflowing gas is potentially crucial to
maintaining the accretion process by carrying away angular momentum;
to chemical evolution in AGN by funneling metal-enriched circum
nuclear gas into their hosts
\citep{1999ASPC..175...33H,2009ApJ...702..851W}; and to regulating
the co-growth of the black holes and the bulges of their host
galaxies by feedback of energy and momentum
\citep[e.g.][]{2002ApJ...574..740T,2000ApJ...539L...9F}, and thus
quenching the gas supply to the nuclear activity and the host star
formation process \citep[e.g.][]{2004ApJ...600..580G,
2005Natur.433..604D}.

AGN outflows often manifest themselves as blueshifted absorption
lines, commonly classified into three categories depending on their
width $\Delta v$ \citep{2004ASPC..311..203H}: (1) broad
absorption-lines (BALs) with $\Delta v\gtrsim 2000~$km~s$^{-1}$; (2)
narrow absorption lines (NALs) with $\Delta v\lesssim 500~$km~s$^{-1}$; (3) mini-BALs with $\Delta v$ in between. NALs are very
common with a detection rate of about 60\% in Seyfert galaxies
\citep{1999ApJ...516..750C} and 50\% in quasars
\citep{2004ASPC..311..203H}. BALs are seen in $\sim 15\%$ of
optically selected quasars\footnote{The frequency depends on the
specific quasar samples and different BAL picking techiques, ranging
from $10\%$ to $40\%$ (e.g. \citet{2002ApJ...578L..31T};
\citet{2003AJ....126.2594R}; \citet{2006ApJS..165....1T};
\citet{2009ApJ...692..758G}; \citet{2008ApJ...672..102G})}. NALs can
originate from gas either intrinsic or intervening to AGN, while
BALs are generally believed to be intrinsic to AGN. All known BAL
quasars show high-ionization broad absorption lines (HiBAL),
including \civ, \siiv\ and \nv. About $15\%$ of BAL quasars present
low-ionization broad absorption-lines (LoBALs) besides HiBALs, such
as \aliii, \alii, and \mgii\
\citep{1991ApJ...373...23W,2003AJ....126.2594R,2010ApJ...714..367Z}.
A minority (another $15\%$) \citep{2002ApJS..141..267H} of LoBAL
quasars exhibit additional absorption in \feii\ and \feiii\ from
both ground and excited levels (FeLoBAL). Even rarer are excited
\hei*\ absorption lines, and Balmer absorption lines are the extreme
rarities.

Only 5 Balmer-absorption AGNs have been reported thus far: NGC 4151
\citep{2002AJ....124.2543H}, SDSS J083942.11+380526.3
\citep{2006ApJ...651...84A}, SDSS J125942.80+121312.6
\citep{2007AJ....133.1271H}, SDSS J102839.11+450009.4
\citep{2008ApJ...674..668W}, and SDSS J172341.10+555340.5
\citep{2010PASJ...62.1333A}. Three of the five are formally
classified as FeLoBALs\footnote{The two non-FeLoBALs are NGC 4151
and SDSS J102839.11+450009.4. The former shows \feii\ NALs
\citep{2002AJ....124.2543H}, and \feii\ NALs are marginally detected
in the later \citep{2008ApJ...674..668W}. }. The high occurrence
rate of both \feii\ and Balmer absorption lines indicates that they
are closely related, in spite of the small number of such AGN. Their
UV spectra are often profuse in excited \feii*\ absorption lines
that are help to constrain the density of absorption gas
\citep{2001ApJ...548..609D}. We would have a more complete picture
of the physical condition and geometry, incorporating ionization
state with \hei* as a useful tracer. Interestingly, both of NGC 4151
\citep{2002AJ....124.2543H} and SDSS J102839.11+450009.4
\citep{2008ApJ...674..668W} show He~I* NALs, which are the only
Balmer-absorption AGN with \hei* covered by previous spectroscopy
with negligible contamination. Note that only a handful of
He~I* absorption AGN have been reported in the
literature\footnote{Previously known He~I* absorption AGN include
Mrk 231 \citep{1977MNRAS.178..451B,1985ApJ...288..531R}, NGC 4151
\citep{2002AJ....124.2543H,1974ApJ...189..195A}, Q 2359-1241
\citep{2001ApJ...546..140A}. \citet{2008ApJ...674..668W} reported
SDSS J102839.11+450009.4 as a remarkable strong narrow \feii\
emission-line quasars with \hei* $\lambda\lambda$3889,~3189,~2796
narrow absorption lines. Recently, \citet{2011ApJ...728...94L}
reported first observation of the infrared line \hei* $\lambda
10830$, which is 23 times stronger than optical \hei* $\lambda 3889$
and could determine a much lower column density.}. The co-occurrence
of He~I* and Balmer absorption lines is again remarkably high.

Balmer absorption lines arise from the absorption by $n=2$ shell
excited hydrogen atoms. H~I* $n=2$ level could be populated by
recombination, collisional excitation and Ly$\alpha$ resonant
scattering \citep{2007AJ....133.1271H,2008ApJ...674..668W}. \hei*
lines arise from absorption by highly metastable level \hei*\ $2^3S$
with transition probability $A(2^3S,1^1S)=1.26\times10^{-4}$~s$^{-1}$.
Transitions from $2^3S$ to different upper levels include \hei*\
$\lambda 10830$ in the infrared \citep{2011ApJ...728...94L},
$\lambda 3889$ in the optical, and $\lambda\lambda 3189,~2946,~2830,~2764$ in the UV. The population of this level comes from the
recombination of He$^+$ ion. A plenty of $>24.6$ eV photons are
needed to ionize neutral helium, while there should not exist too
many $>4.8$ eV photons to ionize \hei*\ $2^3S$. Diffuse stellar
background provides too many photons to ionize the \hei*\ $2^3S$
level, and thus these lines are not easily seen in the interstellar
medium. We can therefore safely associate the incidence of neutral
helium line with an AGN environment \citep{1985ApJ...288..531R}.
Ignoring second-order effects, recombinations of He$^+$ to \hei*\
$2^3S$ is mainly balanced with collisional de-excitation of this
level \citep{2001ApJ...546..140A}. Upon that, He~I* lines can be
used to set a lower limit on the He$^+$ column density, or a lower
limit to \hii\ column density with an assumed abundance.

Potentially we could have a good chance to explore the properties of
AGN Balmer absorption line region (ALR). Due to their rarity,
however, it is not yet clear whether AGN with Balmer ALR form an
unusual class, or they are otherwise normal AGN with our
line-of-sight (LOS) happened to penetrate Balmer ALR, which exists
in most if not all AGN, but with an extremely small global covering
factor (GCF). It is not even known how rare Balmer-absorption AGN
are. Discovery of more such AGN are obviously helpful for
understanding their intriguing nature.

In this paper, we report the new identification of Balmer BALs in
LBQS 1206+1052, which was identified during the large bright quasar
survey \citep{1995AJ....109.1498H} and recently classified as a LoBAL
\citep{2009ApJ...692..758G}. It is in fact one of $\sim 20$
Balmer-absorption quasars we recently identified during our
systematic search in the quasar catalog from the 7th data release
\citep{2010AJ....139.2360S} of Sloan Digital Sky Survey (SDSS,
\citep{2000AJ....120.1579Y}). With $i$ band PSF magnitude of 16.50,
the high brightness of LBQS 1206+1052 makes it an excellent
candidate for high resolution spectroscopic follow-up and for
long-term variability campaign which are crucial for absorption line
study. The paper is organized as follows. In \S~2, we analyze the
SDSS spectrum and measure column densities of H~I*, He~I*, and
Mg~II. The results are discussed in \S~3. \S~4 is a brief summary of
our main conclusions.

\section{SDSS Spectroscopy}

\subsection{Continuum and Emission Line Spectrum}

\begin{figure}
  \centering
   \includegraphics{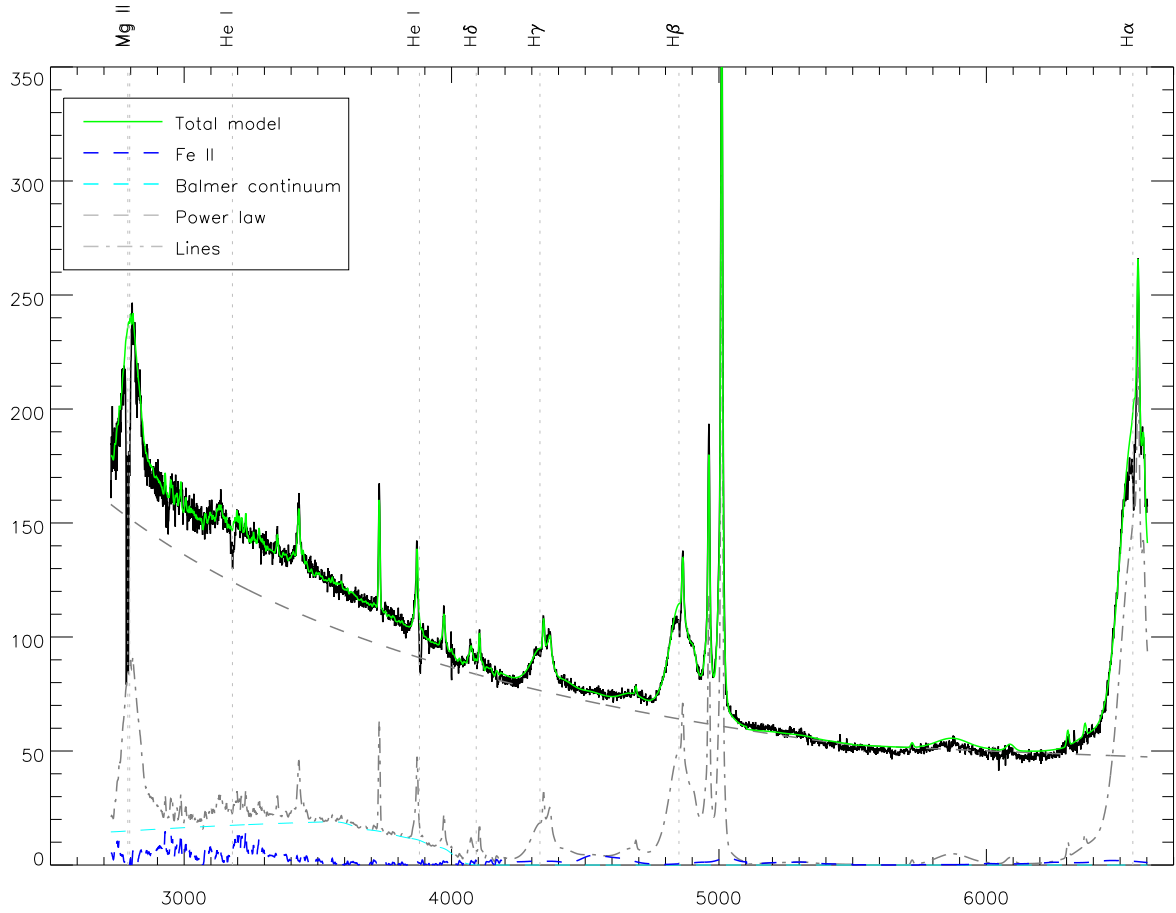}
  \caption[angle=90]{SDSS spectrum of LBQS 1206+1052. Prominent absorption
    features are marked as vertical dashed line and labeled on
    the top. Model fits to the spectra are shown as green line.
Different components of the spectral model are also shown.}
 \label{fig:overview1}
\end{figure}

LBQS 1206+1052 was targeted as a quasar candidate for spectroscopy,
and was observed by SDSS 2.5m telescope on 2003-03-24 with an
exposure time of 3165 seconds. It is very bright with optical PSF
magnitudes of 17.19, 16.63, 16.50, 16.34, 15.92 at $u$, $g$, $r$,
$i$, $z$ bands. The SDSS spectrum, retrieved from SDSS Data Release
7 database, is corrected for a Galactic reddening of $E(B-V)=0.023$
mag \citep{1998ApJ...500..525S}, and brought to the source
rest-frame for further analysis with the systemic redshift
($z=0.3947\pm0.0008$), re-measured from [O {\sc III}] $\lambda 5007$
as described below. The spectrum is shown in Figure
\ref{fig:overview1}. In the rest wavelength range of $\lambda \sim
2500-6600$ \AA, BALs are evident in H$\alpha$, H$\beta$,
\hei*\ $\lambda \lambda$3889,~3189,~2946, and \mgii\ $\lambda
\lambda$2796,~2803 with a velocity range of $\Delta v \sim
0-2000$km~s$^{-1}$.

Properly modeling of continuum and emission lines is essential to
derive absorption line parameters. In the observed quasar spectrum,
prominent broad and narrow emission lines are siting on top of
pseudo-continuum, consisting of thermal emission from the accretion
disk, Balmer continuum from broad line region or disk, the
blended FeII lines and high order Balmer lines. We fit the
pseudo-continuum and emission lines in two steps as follows.  In the
first step, the pseudo-continuum is modeled with three components:
a single power law to mimic emission from accretion disk, a Balmer
continuum plus high order Balmer lines, and blended \feii\ emission
lines.

Following \citet{2003ApJ...596..817D} and
\citet{2006ApJ...650...57T} , the Balmer continuum is expressed by
$F_{\lambda}=F_{3646}\times B_{\lambda}(T_e)(1-e^{-\tau_\lambda})$
for $\lambda \le 3646$~\AA, where $F_{3646}$ is the normalization
coefficient at Balmer edge $3646$~\AA~and
$\tau_\lambda=\tau_{\rm BE}(\lambda/\lambda_{\rm BE})^3$ in which
$\tau_{\rm BE}$ is optical depth at Balmer edge $ \lambda_{\rm BE}$
($3646$~\AA), and $B_\lambda(T_e)$ is the Planck function at an
electron temperature $T_e$, which is assumed to be $15,000$~K
\citep{2003ApJ...596..817D}. High order Balmer lines up to $n=50$ are
also used in the fit. Relative strengths of these lines  are fixed
using the line emissivities for Case~B, $T_e=15,000$~K and
$n_e=10^8$~cm$^{-3}$ as calculated by \citet{1995MNRAS.272...41S},
and each line is assumed a Gaussian profile with a FWHM of 3000 km\
s$^{-1}$.  The relative flux of the high order Balmer lines to the
flux of the Balmer continuum at the edge is fixed using the results
in \citet{1985ApJ...288...94W}. Thus these blended high order Balmer
lines and the Balmer continuum join smoothly, forming a
pseudo-continuum \citep{1985ApJ...288...94W}.  Note that none of the
assumptions concerning the Balmer continuum and high order Balmer
lines will have significant effect on the measurement of absorption
lines because all these features are quite broad.

The optical and UV \feii\ emission are modeled separately with
broadened empirical templates derived from observed quasar spectra.
For optical band long-ward of $3500$~\AA, we use a template extracted
from the optical spectrum of proto-type narrow line Seyfert 1 galaxy
I Zw 1 \citet {2004A&A...417..515V}. The template includes both the
broad and narrow component, the latter also includes transitions
from Ni II, Cr II and Ti II. For UV portion short-ward of $3500$~\AA,
we employ a template built from SDSS J1632+3405 (Zhou et al. 2012 in
preparation). SDSS J1632+3405 has ultra strong UV \feii\  emission,
which has a FWHM of $400$ km\ s$^{-1}$, only about half of that in I
Zw 1. The narrow width makes it easy to separate \feii\  multiplets
from the other lines and also among different multiplets. Thus it is
a more promising target for building UV \feii\  template. The
relative strength of different UV multiplets ratios are allowed to
vary during the fit. This is a common technique when modelling
quasar \feii\ spectra
\citep{2006ApJ...650...57T,2010ApJS..189...15K,2011MNRAS.410.1018S}.

Above continuum components are combined to fit the SDSS spectrum in
following windows: 2855-3010, 3625-3645, 4170-4260, 4430-4770,
5080-5550, 6050-6200~\AA, which are devoid of strong emission lines
\citep{2001ApJS..134....1V}. The window 3625-3645 is used
to constrain Balmer continuum. The best fitted parameters are derived by
minimization of $\chi^2$. Line spectrum is then acquired by subtracting
the best fitted pseudo-continuum from the original spectrum.

In the second step, emission-lines are measured in this line
spectrum. All narrow lines or narrow component of broad lines except
[O {\sc III}] and [Ne {\sc III}] are fitted with a single Gaussian.
The centroids of the Gaussian are fixed to the vacuum wavelengths of
these lines, and their widths are tied during the fit.  [O {\sc
III}] and [Ne {\sc III}] show an additional blueshifted component,
and are fitted with two Gaussians for each of the multiplet. The
width and velocity shift between the two Gaussians are tied for
these lines during the fit, while their normalizations are allowed
to vary independently. [Ne {\sc III}] $\lambda$3869 emission line is affected
by \hei*\ $\lambda$3889 absorption, and the pixels affected by \hei*\ $\lambda$3889
line are masked during the fit according to corresponding \hei*\ $\lambda$3189
absorption profile. Narrow lines included in this fit are [Ne {\sc
V}]~$\lambda$3425, [O {\sc II}], [Ne {\sc III}]~$\lambda$3869,
H$\delta$, H$\gamma$, [O {\sc III}]\ $\lambda$4363, \heii\
$\lambda$4686, H$\beta$, [O {\sc III}]$\ \lambda\lambda$4959, 5007,
[O {\sc I}]$\ \lambda\lambda6300,~6364$,  H$\alpha$, and [N {\sc
II}]$\ \lambda\lambda$6548,~6583.

Broad lines are fitted using as many Gaussians as statistically
justified. Broad component of H$\beta$ is fitted with 4 Gaussians,
while H$\alpha$ with 3 Gaussians because the red part of H$\alpha$
line is not in the observed spectrum range. Pixels affected by
Balmer absorption are masked  as done  for \hei*\ $\lambda$3889. The fitted
profiles of broad H$\alpha$ and H$\beta$ lines are very similar in
the overlapping portion in the velocity space, lending credit to our
fitting scheme of broad lines. \mgii\ lines are fitted with two
Gaussians in the same way as for Balmer lines except that no narrow
Gaussian is used because there are no explicit narrow \mgii\
doublet.

The best fit model is overlaid in Figure \ref{fig:overview1},
together with individual components.

\subsection{Absorption Line Spectrum and Column Densities}

We use the best fit model to normalize the absorption line spectrum.
We first subtract model narrow emission lines from the observed
spectrum before normalization, since absorption gas does not cover
the narrow emission line region (NLR) in all well studied BAL
quasars. \mgii\ doublet and Balmer absorption lines are superposed
on top of broad emission lines. They should be normalized in
different ways depending on whether the absorbing gas covers the
broad emission line regions (BLR) or not
\citep{2001ApJ...548..609D}: (1) The observed data should be divided
by the sum of the continuum and broad emission lines if it does. (2)
If not, the broad emission lines should be subtracted before the
observed data normalized by the continuum. We adopt the second
normalization scheme for the following hints. We bloat Figure 1 and
display the close-ups in Figure 2 focusing on the two H$\alpha$ and
Mg~II regimes. It can be seen clearly that the modeled broad
emission lines just brush the tip of the BAL trough in both of
H$\alpha$ and Mg~II lines. The fact can be most naturally
interpreted in the second scheme with an optical depth $\tau \gg 1$
at the centroid of both H$\alpha$ and Mg~II BAL troughs. Otherwise
we would need to have two coincidences at once, which is very
unlikely if not impossible. The interpretation is also consistent
with the marginal detection of Ca~II~K in absorption, which suggests
a large optical depth in Mg~II BAL doublet for any reasonable
abundance.

The normalized absorption line spectrum is shown in Figure
\ref{fig:normal_second}, and the apparent optical depth profile of \mgii\
$\lambda\lambda$2796,~2803 doublets is shown in Figure \ref{fig:tao}.
The normalized flux at the line centroid approaches zero for \mgii\
and H$\alpha$\ lines. Note that Balmer emission line profiles at the
absorption line position is somewhat uncertain due to lack of
independent constraint. We use 1-$\sigma$ error of normalized flux
to deduce a lower limit for apparent optical depth of the deepest
pixels. The \mgii\ profile apparently show two components, one broad
component with corresponding Balmer and helium absorption lines. The
other one is relatively narrow and shows \hei*\ lines, possible
\caii\ $\lambda$3934 absorption line and little Balmer absorption line. We
simultaneously fit the apparent optical depth $\tau(v)$ of all
absorption lines with two Lorentzians for each line, which is
similar to Voigt profiles for small $b$ values. The line centroid of
other lines are tied to that of \mgii\ lines and the line width is
assumed to be equal for the same component during the fit. The best
fitted model for \mgii\ is also shown in Figure \ref{fig:tao}, and
overlaid in Figure \ref{fig:normal_second}. The fitted EWs for each
line are shown in Table \ref{tab:para}. The low velocity component
is centred at $-726$ km\ s$^{-1}$ and with a FWHM of $603$ km\
s$^{-1}$, it is referred to as component 1 (C1) hereafter, while a high
velocity component with a line centroid of $-1412$ km\ s$^{-1}$ and
a FWHM of $208$ km\ s$^{-1}$ are referred to as component 2 (C2)
hereafter.

The C1 component shows similar profiles in \mgii, \hei*\ and Balmer
absorption lines. There is only one pixel in H$\alpha$ line that is
inconsistent with our model, which can be attributed to imperfect
emission line modeling. The $\tau$ ratio of 2796 to 2803 for C1 is
2:1, which is consistent with that the absorption gas fully cover
the continuum source. To check this,  we compare the ratios of
Balmer lines with their theoretical values. From Table
\ref{tab:para}, we can see that the Balmer absorption line EW ratio
$\rm EW({\rm H}\alpha):\rm EW({\rm H}\beta)=8.19^{+1.29}_{-1.17}$ is consistent with
theoretical $f\lambda$ ratio 7.26, thus we favour that the absorbing
gas of C1 fully covers the continuum source and the lines are not
saturated. Under this assumption, we derive column densities of
various ions from their equivalent widths $W_\lambda$, using the equation
from \citet{1986ApJ...304..739J}:

\begin{equation}
  N=\frac{m_ec^2}{\pi e^2 f \lambda^2} W_\lambda
  \label{columndensity}
\end{equation}

We get a column density for \mgii\ from \mgii\ $\lambda$2803 line:
$N_{\rm Mg^+}= (1.41\pm0.07)\times10^{14}$\ cm$^{-2}$, for hydrogen at
$n=2$ from H$\beta$ line: $N_{\rm H}(n=2)=(1.37\pm0.10)\times10^{14}$\
cm$^{-2}$ and for helium from \hei* $\lambda$3889 line
$N_{{\rm He~I~2^3}S}=(8.99\pm0.50) \times10^{14}$\ cm$^{-2}$

Component C2 shows consistent absorption trough in \mgii\ and \hei*.
A weak line near 3930 \AA\ may be attributed to the \caii\ K
absorption line of this component (blueshifted by $100$ km\ s$^{-1}$
relative to other lines). The \mgii\ absorption line is seriously
saturated as indicated by the doublet ratio of \mgii, ${\rm EW}_{2796}/
{\rm EW}_{2803}\sim1$. Thus we can only get a lower limit of \mgii\ column
density using equation (\ref{columndensity}),
$N_{\rm Mg^+}>(3.43\pm0.34)\times\;10^{14}$\ cm$^{-2}$. Alternatively,
we may integrate the apparent optical depth profile
$N=\frac{m_ec^2}{\pi e^2 f\lambda^2}\int\tau_v~dv$ and get a column
density of $N_{\rm Mg^+}=(5.84\pm0.45)\times{}10^{15}~$cm$^{-2}$, which is
likely close to the real value. The H$\alpha$ line of C2 is not
detected significantly. We derive an upper limit on its column
density $N_{\rm H^0}(n=2) <9\times{}10^{12}$~cm$^{-2}$. The \hei*\
lines in this component are generally weak, we get a column density
of $N_{{\rm He~I~2^3}S}=(1.25\pm0.22) \times{}10^{14}$~cm$^{-2}$.

\section{Discussion\label{sec:discussion}}

The overall rarity of Balmer absorption line quasar indicates that
either the conditions for outflowing gas to maintain a sufficient
population of $n=2$ level hydrogen atoms must be very harsh. We will
discuss the conditions required to produce the Balmer absorption
lines in C1 with different mechanisms.

If the $n=2$ level hydrogen atoms of this component are caused by
collisional excitation, they must exist in \hi\ region or probably
the partially ionized region. The observed \mgii\ absorption lines
suggest that there is partially ionized absorption line region.
Mg$^+$ ions are created by photons with energy greater than 7.6 eV
and destroyed by photons with energy greater than 15.0 eV, thus they
mainly survive in outer part of \hii\  region and mostly in
partially ionized region. Assuming that Mg$^+$ is the dominant
species of Mg and a solar abundance with [Mg/H]=-4.4, we got a lower
limit on the total hydrogen column density of $3.54\times10^{18}$\
cm$^{-2}$ for the partially ionized zone.  If thermal equilibrium is
achieved, then using Boltzmann equation: $n_2/n_1=4
{\rm exp}(-10.2$eV$/kT)$, in which $ n_1$ and $ n_2 $ are densities of
$n=1$ and $n=2$ level hydrogen, one yields readily a column density
of $n=2$ hydrogen of $6.8\times10^{13}$\ cm$^{-2}$, which is close
to the observed hydrogen column density $(1.37\pm 0.10)\times
10^{14}$\ cm$^{-2}$, at a typical temperature T$\sim 10^4$K for
photo-ionized gas. However, thermal equilibrium requires that
collisions with thermal particles are the dominant process for
excitation and de-excitation. This yields a lower density limit
$n_c=A_{21}/[q_{21}(1+\tau_{0l})]=8.7\times10^{16}/(1+\tau_{0l})$
cm$^{-3}$ accounting for the resonant scattering of Ly$\alpha$
photons, where $\tau_{0l}$ is the optical depth of Ly$\alpha$ from
the central of the region.  For a gaussian distribution turbulence,
we can write $\tau_{0l}\simeq 7.6\times 10^6b_3^{-1}N_{\rm H^0,22} \sim
10^3$, where $b_3\sim 1$ is the turbulent velocity of gas in units
of $10^3$ km~s$^{-1}$ and $N_{\rm H^0,22}$ neutral hydrogen column
density in $10^{22}$ cm$^{-2}$. This gives a density too high for
any reasonable line absorbing gas.

When density is lower than the critical density discussed above, detailed
equilibrium of $n=2$ level must be considered. Transition of $2s$ state to
the ground level is forbidden with a small Einstein coefficient $A=8.23~s^{-1}$
via 2-photon process, while transition from $2p$ to $1s$ is allowed. The two
levels have very different excitation processes, therefore, we will consider
them separately in the following analysis.

Let us first consider the equilibrium of $2p$ level. In the case
that absorption gas is optically thin to Ly$\alpha$, the equilibrium
between the recombination to $n=2$ and radiative decay from $2p$ to
$2s$ yields $n_{2p}/n_{\rm H^+}=\alpha_{\rm eff} n_e/A_{2p,2s}$\footnote{As
we discuss next, the absorbing gas is probably in partially ionized
region, where the typical temperature is significantly lower than
$10^4$~K, and $q_{1s-2p}\propto {\rm exp}(-\frac{\Delta
E}{kT})\ll{}\alpha_{\rm eff}$ in this region, thus we ignore the
collisional excitation of 2p level here.}. However, the gas is
likely very optically thick to Ly$\alpha$. In this case, each
Ly$\alpha$ photon is scattered $1+\tau_{0l}$ times on average before
escaping \citep{2006agna.book.....O}. This leads to the $2p$
population larger by a factor of $1+\tau_{0l}$ than above optically
thin formula, i.e., $n_{2p}/n_{\rm H^+}\simeq\tau_{\rm 0l}\alpha_{\rm eff}
n_e/A_{2p,2s}\simeq 7.3\times10^{-9}n_{e,6}N_{\rm H^0,22}b_3^{-1}$. To
get the observed $n=2$ column density, we require
\begin{equation}
\label{eq:H_column_ionization} N_{\rm H,22}\simeq 1.4 f_{\rm H^+}^{-1}
(1-f_{\rm H^+})^{-1/2}b_3^{1/2}n_{\rm H,6}^{-1/2} (N_{n=2}/1.4\times 10^{14}
~{\rm cm}^{-2})^{1/2},
\end{equation}
where $f_{\rm H^+}$ is fraction of ionized hydrogen. As discussed in
\S 2.2, it is very likely that the BALR in LBQS 1206+1052 fully
covers the accretion disk but does not cover the BLR. This suggest
$10^{14}~{\rm cm} \sim r_{\rm AD}<r_{\rm BALR}\ll r_{\rm BLR} \sim 10^{17}$~cm, where
$r_{\rm AD}$, $r_{\rm BALR}$ and $r_{\rm BLR}$ are the size of accretion disk,
broad absorption line region, and broad emission line region,
respectively. We infer a hydrogen column density of $N_{\rm H}\sim
10^{21-22}$ cm$^{-2}$ of the partially ionized gas for an electron
density of $n_{\rm H}\sim 10^{6-8}~$cm$^{-3}$. For a higher $n_{\rm H}$ and lower
$N_{\rm H}$, the BALR would not be large to fully cover the accretion
disk, while the BALR would be large enough to cover the BLR for a
lower $n_{\rm H}$ and higher $N_{\rm H}$. The inferred $N_{\rm H}$ value is much
larger than the thermal equilibrium value.

Next, we consider population in the $2s$ level. The equilibrium
equation of $2s$ level can be written as
\begin{equation}
\label{eq:H_equilibrium}
 n_en_{\rm H^+}\alpha_{{\rm eff},2s}+ n_{\rm H}n_eq_{1s-2s}=
n_{2s}\left[A_{21}+n_e(q_{2s-2p}+q_{2s-nl})+\int^\infty_{\nu_1}\frac{\alpha_\nu
L_\nu}{4\pi r^2h\nu}d\nu\right],
\end{equation}
in which we ignored the effect of $2p$ level for the time
being\footnote{Had $2p$ been significantly populated due to
Ly$\alpha$ trapping in dense gas as discussed in last paragraph,
because of large collision strength, $2p$ and $2s$ level will reach
thermal equilibrium, that is $n_{2s}:n_{2p}=1:3$.}. The first term
on the left hand represents recombination to $2s$ level, while the
second term represents collisional excitation. Given that
$\alpha_{{\rm eff},2s}$ and $q_{1s-2s}$ are in the same order of magnitude
($\alpha_{{\rm eff},2s}=0.838\times10^{-13}$\ cm$^{-3}$\ s$^{-1}$,
$q_{1s-2s}=1.67\times10^{-13}$~cm$^{-3}$~s$^{-1}$ at T=10,000 K),
the ratio of the first to second term is mainly determined by the
ratio of $n_{\rm H^+}$ to $n_{\rm H}$. In \hii\ region where
$n_{\rm H^+}/n_{\rm H}$ is high, the recombination mechanism dominates the
population of $n=2$ level, while in \hi\ region or PIZ where
$n_{\rm H^+}/n_{\rm H}$ is low, the collisional excitation dominates the
population of $n=2$ level, we treat them separately.

In \hi\ or PIZ region, we ignore the first term on the left hand of
equation (\ref{eq:H_equilibrium}). The third term on the right hand
is the photoionization of $n=2$ level, in which $\nu_1$ is the
frequency at Balmer edge  $\lambda=3646$~\AA. It can be shown that
the third term is negligible by substituting the integration term
with $U=\int^\infty_{\nu_0}L_\nu/(4\pi r^2c n_{\rm H}h\nu)d\nu$ and
$\nu_0=4\nu_1$ and assuming a power low index of optical-UV
continuum $\alpha_\nu=-0.44$ \citep{2001AJ....122..549V}. If density
$n_e$ of absorption gas is sufficiently low compared to critical
density $n_c=A_{2S\rightarrow 1S}/q_{2S\rightarrow 2P}\sim 1.5\times
10^4$\ cm$^{-3}$ at $T=10,000$~K (Osterbrock\& Ferland 2006), the
equilibrium gives
$n_{2S}/n_{\rm H}=n_eq_{1S\rightarrow2S}/A_{2S\rightarrow1S} \leq
q_{1S\rightarrow2S}/q_{2S\rightarrow2P}$, otherwise if the density
is moderately high, the second route will dominate, which gives
$n_{2S}/n_{e}=q_{1S\rightarrow2S}/q_{2S\rightarrow2P}$, which is
around $3\times{}10^{-10}$ at a temperature of $10,000$~K
\footnote{\citet{2008ApJ...674..668W} used an incorrect
  $q_{2S\rightarrow2P}$  value resulting an incorrect value
  $n_{2S}/n_{\rm H}\sim  0.02$}.
This requires a total hydrogen column greater than $5\times10^{23}$
cm$^{-2}$ to generate the observed Balmer absorption lines EW in Q
1206+1052. Clearly, collisional excitation is mush less efficient in
populating $n=2$ level than Ly$\alpha$ resonant scattering in
partially ionized gas in AGNs if gas density is high.

Alternatively the $n=2$ level population could arise from ionized gas via
recombination, and we ignore the second term on the left hand of equation
(\ref{eq:H_equilibrium}). As in the collisional excitation dominated case,
we further distinguish cases when densities are below or above the critical
density $10^4$\ cm$^{-3}$ set by collisional angular-momentum transition
from $2s$ to $2p$ state.  If the density is higher than that, the $2s$ state
is populated by recombination and depopulated via collisional excitation
to $2p$ state, which radiatively decay to $n=1$ level. The equilibrium gives
$n_{2s}/n_{\rm H^+}\sim\alpha_{2s}/q_{2s\rightarrow2p}\sim 0.44\times 10^{-10}$
at $T=10,000$~K.  Otherwise if the density is lower than that critical
density, the equilibrium is built through the balance between the
two-photon continuum emission of $2s\rightarrow1s$ and the
recombination to 2s state, which gives $n_{2s}/n_{H+}\sim
n_e\alpha_{2s}/A_{2s\rightarrow1s}\le\alpha_{2s}/q_{2s\rightarrow2p}\sim0.44\times10^{-10}$
. The last equation in above estimations utilizes the fact that when this
route dominates the equilibrium, $n_e\le n_c=A_{2s\rightarrow1s}/
q_{2s\rightarrow2p}$.  Thus if indeed the level two population is from
the HII zone, the observed  $n=2$ column require a total hydrogen column
of at least $3.1\times10^{24}~$cm$^{-2}$. The column density appears too
large for the absorbing gas.

To summarize, the resonant scattering of Ly$\alpha$ is much more efficient
in populating $n=2$ level in the partially ionized region if gas density is
high. In this case, the column density required to account for the Balmer
absorption lines in C1 are in the range of $10^{21}$ to $10^{22}$~cm$^{-2}$
for gas density between $10^6$ to $10^8$ cm$^{-3}$ and decreases with square
roots of the gas density. At much low densities, collisional excitation may
take over the resonant scattering pumping, and it requires a column density
at least 5$\times10^{23}$~cm$^{-2}$. In the \hii\ region, recombination can
lead to the required $n=2$ level at very high column density $N_{\rm H^+}=
5\times 10^{24}$~cm$^{-2}$.

A thick \hii\ region is not consistent with the observed helium
absorption lines. \hei*\ absorption arises from absorption by
metastable \hei*\ $2^3S$ state. The equilibrium equation of this
level is \citep{2001ApJ...546..140A}:
\begin{equation}
  n_{\rm
    He+}n_e\alpha_T=n_{2^3S}\left[A_{21}+n_e(q_{2^3S,2^1S}+q_{2^3S,2^1P})+n_eq_{\rm
      ci}
    +\int^\infty_{\nu_0}\frac{\alpha_\nu L_\nu}{4\pi r^2h\nu}d\nu\right]
\end{equation}
where, $\nu_0$ is the threshold frequency at ionization energy
$4.77$eV of $2^3S$ state, and $q_{2^3S,2^1S}$, $q_{2^3S,2^1P}$,
$q_{\rm ci}$ are corresponding collisional rates. The left hand term
represent recombination of \heii\ ions with electrons. On the right
hand of the equation, the first term is radiative transition from
$2^3S$ to $1^1S$ , which is a forbidden transition and can be
neglected; the second term are collisional de-excitation of $2^3S$
to $2^1S$ and $2^1P$; and the third and forth terms represent
collisional and radiative ionization of $2^3S$ respectively.
Neglecting terms other than collisional de-excitation, we got
\begin{equation}
  \label{eq:arav}
  \frac{n_{2^3S}(\rm He^0)}{n_{\rm He^+}}=\frac{5.8\times 10^{-6}T_4^{-1.19}}{1+3110T_4^{-0.51}n_e^{-1}}
\end{equation}
where $T_4=T/10^4$~K~$\simeq 1$ for photoionized gas. At a reasonable gas density,
the observed \hei* $2^3$S column density implies a column density of \hii\ zone of
only $10^{21}$ cm$^{-2}$. This column density is much lower than
required for excitation of $n=2$ level via recombination, but is consistent
with Ly$\alpha$ pumping as the main excitation mechanism for $n=2$ population.

A thick \hi\ region can produce too larger column density of \mgii. It is
shown that the fraction of Mg$^+$ in the PIZ of AGN ionized region ranges
from $10^{-3}$~cm$^{-3}$ close to \hii-\hi\ transition zone to near dominated species
deep inside the \hi\ region. To meet the observed \mgii\ column density,
the PIZ zone can not be thicker than a few $10^{21}$~cm$^{-2}$, which
 rules out the possibility of collisional excitation without Ly$\alpha$
pumping. Therefore, Ly$\alpha$ pumping remains the only possible excitation
mechanism at least for the Balmer absorption in this object, and the
density is likely higher than $10^8$~cm$^{-3}$.

For C2, \mgii\ lines are seriously saturated. The column densities
derived from integrated \mgii\ apparent optical depth profile set a
lower limit on hydrogen column density of
$N_{\rm H}=6.23\times10^{19}~$cm$^{-2}$. On the other hand, a lower limit of
the column density ($N_{\rm H}\sim 10^{21}$~cm$^{-2}$) can be derived from
the \hei* absorption lines with an analysis similar to that for
C2. Although \mgii\ column density is much larger than that of C1
and \hei* column density is similar, the Balmer absorption is
not detected with an upper limit of $9\times10^{12}$~cm$^{-2}$.

As discussed above, $n=2$ hydrogen is populated much more efficiently
via Ly$\alpha$ pumping process than with other routes. The process
works well in relatively dense material and in partially ionized
medium. Lack of the Balmer absorption lines can be attributed to any
of the following cases (\ref{eq:H_column_ionization}): the total
column density of partially ionized zone is too low; the ionization
of the gas may be too large; the density of gas is too low, or any
combination of the three. As pointed out in
\citet{2008ApJ...680..858L}, a trace amount of Mg$^+$ may originate
from an ionized region with log$~U\geq -1.5$ where Al$^{2+}$ and
C$^{3+}$ live.  A total hydrogen column density of
$N_{\rm H}\sim10^{22}$~cm$^{-2}$ is required to generate the observed Mg$^+$
if this is the case, although the column is much larger than the
lower limit set by helium absorption, we can not rule out this
possibility for now.

\section{Summary and Future Prospectives}

We present a detailed analysis of absorption line systems in the
quasar LBQS 1206+1052, the brightest quasar showing Balmer
absorption lines H$\alpha$ and H$\beta$. From its medium resolution
SDSS spectra, we found that there are mainly two absorption
components. Component 1 is blueshifted by $-726$~km~s$^{-1}$ and
with a FWHM of 603 km~s$^{-1}$,  showing corresponding absorption
in Balmer, \hei*\ and \mgii. This component has full coverage of the
continuum source but not BLR. Component 2 is blueshifted by -1412
km\ s$^{-1}$ and with a FWHM of $207$~km~s$^{-1}$. The \mgii\ lines
are seriously saturated for this component, while helium lines are
generally too weak to facilitate partial coverage analysis and
Balmer absorption lines are not detected. If the H$(n=2)$ is the
outcome of recombination, a typical ionized hydrogen column density
of several $10^{24}$~cm$^{-2}$ is needed to generate the observed
absorption in Balmer series. Collisional excitation of 2$s$ level
without Ly$\alpha$ resonance scattering in play also requires an
column as high as $5\times 10^{23}$~cm$^{-2}$. These column density are
inconsistent with either \mgii\ or \hei$^*$\ absorption lines.
Thus, Ly$\alpha$ scattering in a partially ionized region is the
main mechanism of populating $n=2$ level hydrogen, and the required
column is $10^{21}$ to $10^{22}$~cm$^{-2}$ for gas density between
$10^6$ to $10^8$~cm$^{-3}$ and decreases with square roots of the
gas density. These harsh conditions required, i.e. the large column
density and density, may help to explain the overall rare population
of such objects, only $\sim20$ (Ji et al. 2011 in preparation) out
of $105783$ DR7 quasars \citep{2010AJ....139.2360S}.

Due to the medium S/N and resolution of SDSS spectra, we cannot
restrain further about the nature of the absorber. However, the
bright magnitudes make LBQS 1206+1052 the most promising candidate
for high resolution spectroscopic observation, which will provide
resolved velocity profile for pixel-to-pixel based covering factor
and velocity structure analysis. The Earth's atmosphere cutoff
wavelength $3200~$\AA~corresponds to a rest-frame wavelength
$2285~$\AA~at $z=0.4$ of the quasar. Given the high rate of
detecting Balmer absorption in FeLoBALs, we expect to detect
absorption lines in UV1 multiplet arising from excited levels of
\feii\ ions from ground-based facilities, such as Keck and MMT,
which may help to determine the density of the absorption gas
\citep{2008ApJ...688..108K}. Long-term monitor is also helpful to
explore its physical/geometrical properties.

This work is supported by the Chinese NSF through grants NSF
10973013 and 11033007, and the Fundamental Research Funds for the
Central Universities through grant WK 2030220006, and the SOA
project CHINARE2012-02-03. This paper has made use of data from
NED,NIST and SDSS. Funding for the SDSS and SDSS-II has been
provided by the Alfred P. Sloan Foundation, the Participating
Institutions, the National Science Foundation, the U.S. Department
of Energy, the National Aeronautics and Space Administration, the
Japanese Monbukagakusho, the Max Planck Society, and the Higher
Education Funding Council for England. The SDSS Web Site is
http://www.sdss.org/. The SDSS is managed by the Astrophysical
Research Consortium for the Participating Institutions. The
Participating Institutions are the American Museum of Natural
History, Astrophysical Institute Potsdam, University of Basel,
University of Cambridge, Case Western Reserve University, University
of Chicago, Drexel University, Fermilab, the Institute for Advanced
Study, the Japan Participation Group, Johns Hopkins University, the
Joint Institute for Nuclear Astrophysics, the Kavli Institute for
Particle Astrophysics and Cosmology, the Korean Scientist Group, the
Chinese Academy of Sciences (LAMOST), Los Alamos National
Laboratory, the Max-Planck-Institute for Astronomy (MPIA), the
Max-Planck-Institute for Astrophysics (MPA), New Mexico State
University, Ohio State University, University of Pittsburgh,
University of Portsmouth, Princeton University, the United States
Naval Observatory, and the University of Washington. \clearpage

\clearpage
\begin{figure}
 \includegraphics{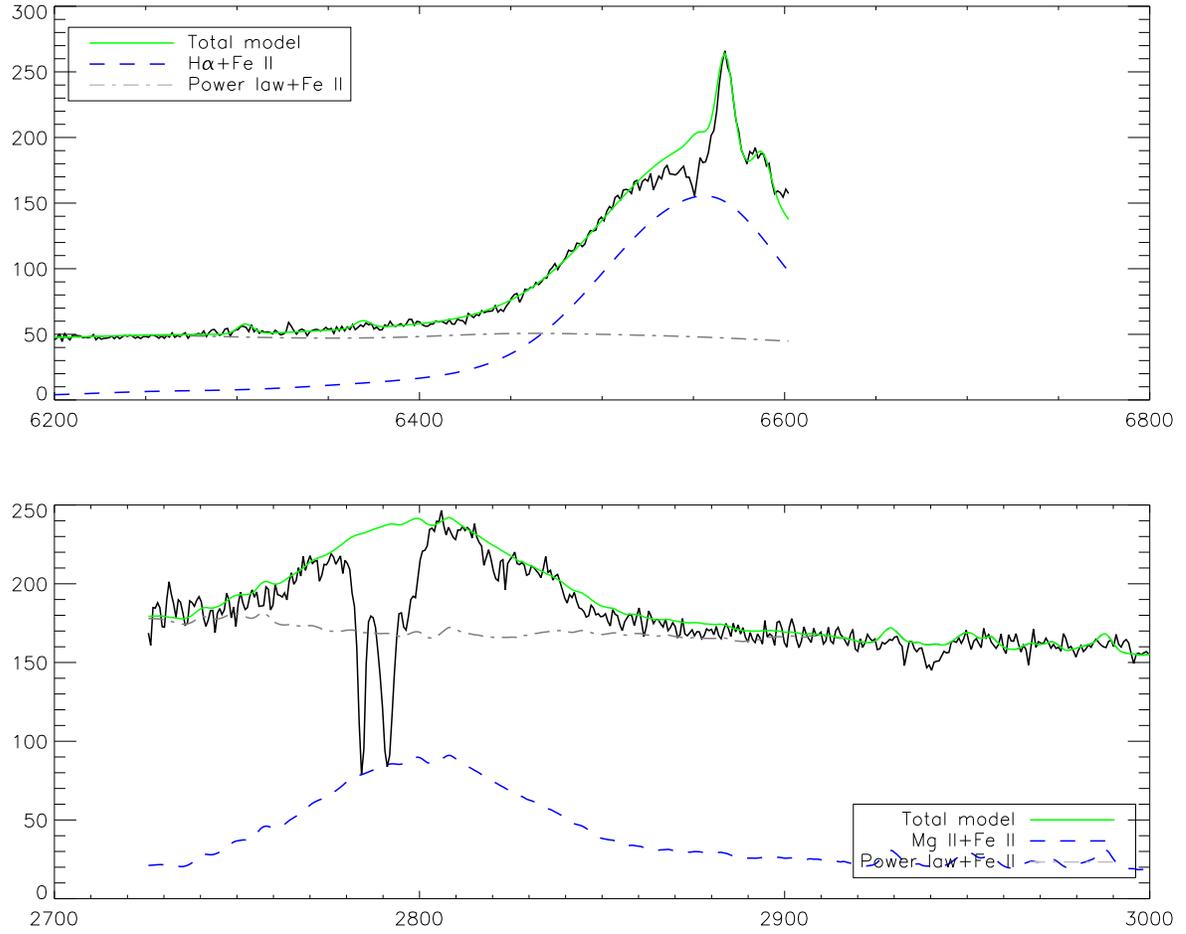}
  \caption{Close-ups of Figure 1 bloated in the H$\alpha$ (upper) and
    Mg II (lower) regimes.}
 \label{fig:normal_first}
\end{figure}

\begin{figure}
\centering
  \includegraphics[scale=0.8]{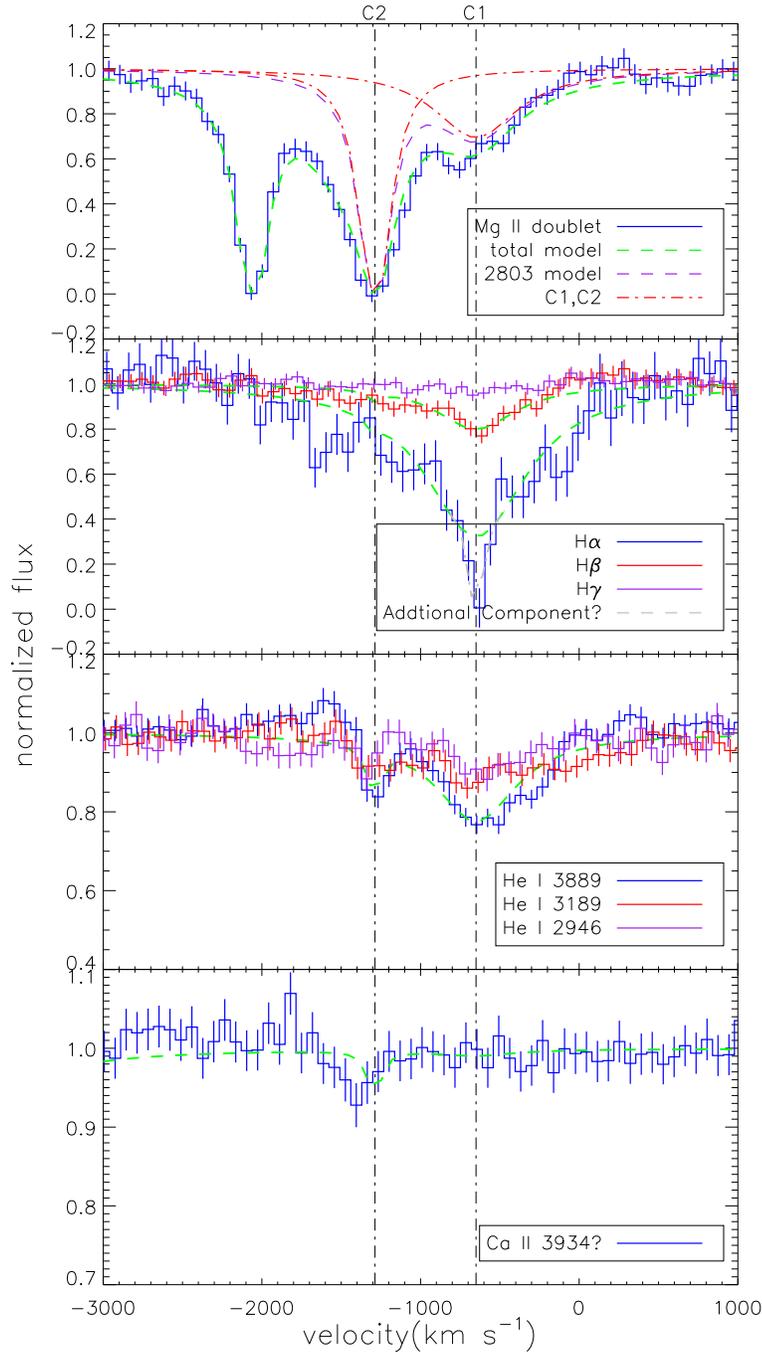}
  \caption{Normalized intensity assuming that absorption gas
    cover only the continuum source but not the BLR. From top to bottom are
    \mgii\ $\lambda$
    2803, Balmer absorption lines and \hei*\ absorption lines,
    \caii\ $\lambda$3934.
    Different lines are plotted using different colors as indicated by legends
    on the lower right of each panel. Best fitted models of absorption profile
    are shown as green lines, only fits for \mgii, H$\alpha$, \hei*\
    $\lambda$3889 and  \caii\ $\lambda$3934 are
    shown for clarity. The two components are shown in red dot-dashed
    lines for \mgii\ fit, whose centres are marked as black vertical
    lines.. The absorption model is oversimplified as
    judged by the central pixels of H$\alpha$ line, yet has little
    effect on our analysis.}
\label{fig:normal_second}
\end{figure}

\begin{figure}

  \includegraphics{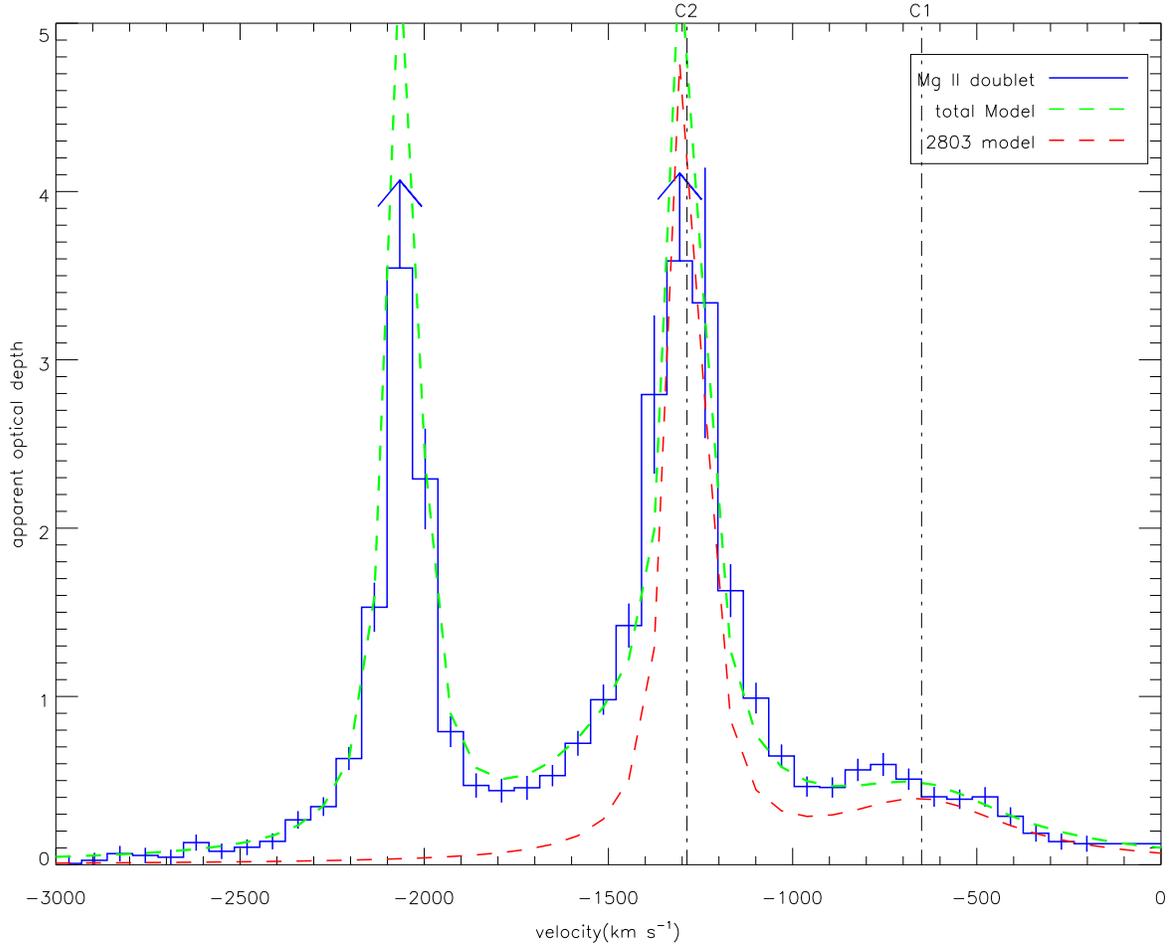}
  \caption{Apparent optical depth for \mgii\ $\lambda \lambda$2796,~2803 doublet as
    function of velocity relative to rest-frame 2803~\AA. The
    pixels marked as up-arrow are lower limits of optical depth
    values. These pixels has normalized flux approaching zero and
    1-$\sigma$ errors of normalized flux are used to get the optical
    depth limit. Various components are also plotted using different
    lines and colors as  indicated by legends.}
\label{fig:tao}
\end{figure}
\clearpage

\begin{table}
\begin{minipage}{\linewidth}
\caption{Absorption line parameters}
\label{tab:para}
\tabcolsep 6mm
\small
\centering
\begin{tabular}{lcccc}
  \hline\noalign{\smallskip}
Line&Wavelength(\AA)&$f_{ij}$\footnote{ Oscillator strength are from
  NIST Atomic Spectra Database
  (http://physics.nist.gov/PhysRefData/ASD/).}&$W_\lambda({\rm C1})$\footnote{Equivalent widths for component 1, which is a
    Gaussian with centroid at -726 km\ s$^{-1}$ and FWHM 603 km\
    s$^{-1}$}&$W_\lambda({\rm C2})$\footnote{Equivalent widths for component 2, which is a
    Gaussian with centroid at -1412 km\ s$^{-1}$ and FWHM 208 km\ s$^{-1}$}\\
\hline\noalign{\smallskip}
  H$\alpha$&6564.41&0.6400&$27.88\pm2.16$&$<0.70$\\
  H$\beta$&4862.68&0.1190&$3.41\pm0.25$&$0.15\pm0.08$\\
  H$\gamma$&4341.68&0.0446&$0.46\pm0.13$&none\\
  \hei*\ 3889&3889.75&0.0278&$3.35\pm0.19$&$0.47\pm0.08$\\
  \hei*\ 3189&3188.66&0.0110&$1.46\pm0.13$&$0.19\pm0.06$\\
  \hei*\ 2946&2945.96&0.0053&$0.91\pm0.13$&$0.15\pm0.06$\\
  \mgii\ 2796&2796.35&0.6160&$6.00\pm0.31$&$7.29\pm0.73$\\
  \mgii\ 2803&2803.53&0.3060&$3.01\pm0.15$&$7.31\pm0.73$\\
  \caii\ 3934&3934.78&0.6270& none& $<0.24$\\
  \noalign{\smallskip}\hline
  \end{tabular}
\end{minipage}
\end{table}

\clearpage
\bibliography{./all_no_abstract}
\end{document}